# TWO-STAGE COORDINATION MULTI-RADIO MULTI-CHANNEL MAC PROTOCOL FOR WIRELESS MESH NETWORKS

Bingxuan Zhao and Shigeru Shimamoto

Graduate School of Global Information and Telecommunication Studies, Waseda University, Tokyo, Japan

zhaobx@fuji.waseda.jp, shima@waseda.jp

## ABSTRACT

*Within the wireless mesh network, a bottleneck problem arises as the number of concurrent traffic flows (NCTF) increases over a single common control channel, as it is for most conventional networks. To alleviate this problem, this paper proposes a two-stage coordination multi-radio multi-channel MAC (TSC-M2MAC) protocol that designates all available channels as both control channels and data channels in a time division manner through a two-stage coordination. At the first stage, a load balancing breadth-first-search-based vertex coloring algorithm for multi-radio conflict graph is proposed to intelligently allocate multiple control channels. At the second stage, a REQ/ACK/RES mechanism is proposed to realize dynamical channel allocation for data transmission. At this stage, the Channel-and-Radio Utilization Structure (CRUS) maintained by each node is able to alleviate the hidden nodes problem; also, the proposed adaptive adjustment algorithm for the Channel Negotiation and Allocation (CNA) sub-interval is able to cope with the variation of NCTF. In addition, we design a power saving mechanism for the TSC-M2MAC to decrease its energy consumption. Simulation results show that the proposed protocol is able to achieve higher throughput and lower end-to-end packet delay than conventional schemes. They also show that the TSC-M2MAC can achieve load balancing, save energy, and remain stable when the network becomes saturated.*

## 1. INTRODUCTION

As an emerging and promising technology, a wireless mesh network (WMN) can provide high QoS to end users as the last mile technique for data delivery over the Internet. Different from mobile ad hoc networks (MANET), WMN has its own unique features [1]-[3]. At first, it has an infrastructure consisting of stationary or slow mobile wireless routers and gateways through which mesh clients connect to the Internet. Obviously, the gateway has much stronger computing capability than other nodes. Secondly, its traffic pattern is mainly from the clients to the gateways or in the inverse direction. Therefore, the links closer to the gateway are more likely to be burdened with a higher traffic load. Therefore, the design of protocol for WMN should take such unique features into full consideration in order to achieve better performance.

Compared to single channel MAC protocols, multi-channel MAC protocols can achieve much higher time-spatial-reuse efficiency and thus considerably improve system performance. The multi-radio multi-channel MAC (M2MAC) [4]-[7] can further improve the flexibility of resource allocation, and it provides at least two more advantages over the single radio multi-channel MAC [8][9]. First, radios do not always need to switch among different channels, which will simplify the design and significantly reduce the protocol overhead. Second, M2MAC can further improve network capacity, because it features nodes that can communicate simultaneously on different radios. Considering the fact that IEEE 802.11b/g and 802.11a can respectively provide as many as 3 and 13 orthogonal channels [10], it makes sense to study M2MAC for practical applications.

To the best of our knowledge, almost all the previous M2MAC protocols [4][6][7][9] specify an out-of-band or in-band single channel as the common control channel (CCC) for exchanging control messages so that the protocol can coordinate the communication pair nodes and mitigate hidden nodes problem. In this paper, we consider only the in-band control channels. Obviously, the use of dedicated CCC could cause resource waste when the Number of available Orthogonal Channels (NOC) is small (e.g., 1/3 orthogonal channels have to be used as a control channel in IEEE 802.11 b/g), so it may face the bottleneck problem when the Number of Concurrent Traffic Flows (NCTF) grows immense. In the latter case, the collision of control messages will lead to system performance degradation in terms of throughput and end-to-end packet delay.

In this paper, we propose a Two-Stage Coordination M2MAC (TSC-M2MAC), which not only solves the abovementioned bottleneck problem for control messages exchanging but also alleviates hidden node problems and achieves load balancing among different channels. The proposed TSC-M2MAC exploits all available orthogonal channels for both control messages exchanging and data transmission through a time division manner. Compared to conventional M2MAC, the proposed TSC-M2MAC has the following new features:

a)   Instead of exploring a single dedicated CCC, all the available channels are designated as control channels and data channels on Channel Negotiation and Allocation (CNA) and Data Transmission (DT) sub-intervals, respectively, in a time division manner. It can also address the scenario of large NCTF.

b)   An intelligent control channel allocation algorithm, which considers co-channel interference, is proposed at the first stage. It is able to achieve load balancing among all channels, and it minimizes co-channel interference.

c)   At the second stage, a REQ/ACK/RES mechanism is proposed to realize dynamical channel allocation for data transmission. The hidden node problem can also be alleviated with channel and radio utilization structures.

d)   An adaptive adjustment algorithm (AAA) for the CNA sub-interval is proposed to cope with the variation of the number of traffic flows at different times.

e)   A Power Saving Mechanism (PSM) designed specifically for the TSC-M2MAC is presented to improve the efficiency of its energy consumption.

The remainder of this paper is organized as follows. Section 2 presents the system model. Section 3 reports the admissible conditions for new wireless links. Sections 4 and 5 show the proposed Control Channel Allocation Algorithm (CCAA) and the dynamic data channel allocation, respectively. We present the AAA for the CNA sub-interval and PSM in sections 6 and 7, respectively. Section 8 describes the simulation-based performance evaluation. We conclude our paper in section 9.

## 2. SYSTEM MODEL OF TSC-M2MAC

TSC-M2MAC is designed for a heterogeneous system environment. There is no requirement for the radio communication or sensing range, the physical layer technology of each radio, or the number of radios equipped on each node. The two main assumptions in TSC-M2MAC include: 1) all the nodes in the network are synchronized, similar to [9]; and 2), all the radios used in the network are half-duplex, which means the radio cannot transmit data and receive data at the same time. The TSC-M2MAC is a virtual MAC running on the top of all radios equipped on each node. Figure 1 demonstrates the schematic of a multi-radio node. This node's objectives are to increase throughput, to decrease end-to-end delay, and to reduce energy consumption.

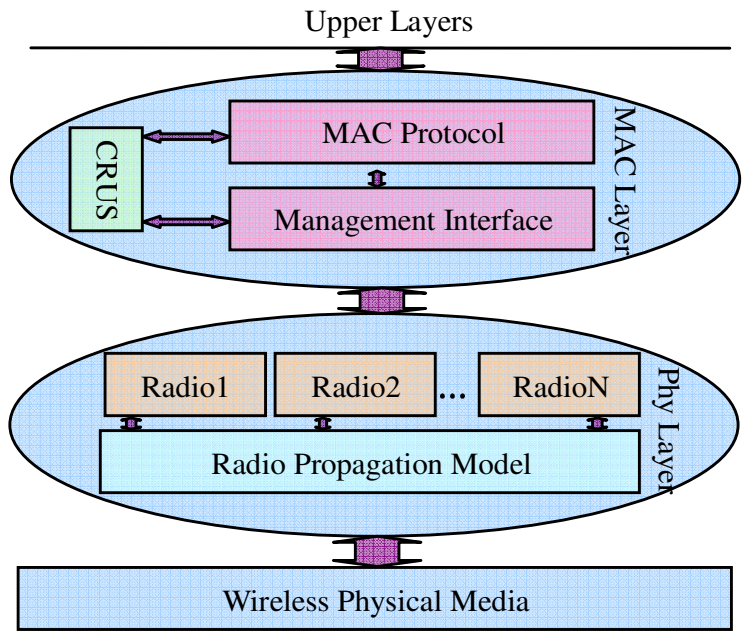

Figure 1.    Schematic of a Multi-Radio Node

To realize such objectives by efficiently utilizing the available channels, TSC-M2MAC consists of a two-stage coordination process: control channel allocation and data channel allocation, which are performed in a centralized and distributed manner, respectively. At the first stage, in order to minimize co-channel interference and to decrease the collision probability of control messages (similar to [6]), we apply the Multi-radio Conflict Graph (MCG) to model the interference, but we use a different BFS-based vertex coloring algorithm (VCA) in order to consider load balancing among all channels. The VCA can figure out control channel allocation results for radios on different nodes. VCA is performed through the gateway and is triggered by topology variations. Given the complexity of VCA and the slow change of traffic patterns, this algorithm can be executed for a long period, e.g., at 100 beacon intervals, so as to alleviate its demand on the gateways. At the second stage, a REQ/ACK/RES mechanism is used to dynamically allocate channels for data transmission. Control messages (i.e., beacon, data pilot, REQ, ACK, RES) are transmitted over allocated control channels at the first stage. Similar to the RTS/CTS mechanism, the REQ/ACK/RES also has an exponential backoff mechanism to avoid collision.

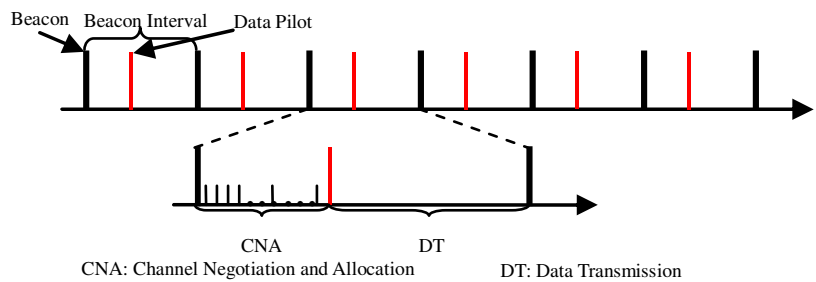

Figure 2.    Time Division Method in TSC-M2MAC

At the second stage in TSC-M2MAC, the timeline is divided into equal beacon intervals. The periodical data pilots further divide each beacon interval into the CNA sub-interval and the DT sub-interval, as shown in Figure 2. The CNA sub-interval is used for channel allocation for data packets. In CNA, control messages like REQ, ACK, and RES are exchanged over the allocated control channels. The REQ/ACK/RES mechanism can allocate every communication radio-pair with an appropriate channel for data packets. We exploit the AAA for the CNA sub-interval to cope with the variation of NCTF. In the DT sub-interval, data packets are transmitted over the negotiated data channels.

## 3. ADMISSIBLE CONDITIONS FOR NEW WIRELESS LINKS

Generally speaking, NCTF in the network is much bigger than NOC. So, it is inevitable that more than one wireless link, i.e., the links between the receiver and the transmitters in the sensing range of the receiver, share one channel. To schedule a new wireless link to share a channel, the schedule algorithm should be satisfied with two conditions. First, the interference caused by the new link should not lead to decoding errors in the existing receiving nodes operating on the channel. Second, the existing interference in the channel should not be strong enough to impact the correct decoding on the receiving node of the new wireless link. The method to determine these two conditions is as follows.

To correctly decode packets at the receiver, $Rx(l_i)$, of the $i^{th}$ link, $l_i$, the received SINR should be greater than a predefined threshold $\gamma$ if we assume that the threshold is the same for all the radios in the network. Such a condition can be expressed by:

$$\frac{P_r(l_i)}{N + \sum_{l_j, j \neq i} I_{Rx(l_i)}^{Tx(l_j)}} > \gamma \Leftrightarrow \sum_{l_j, j \neq i} I_{Rx(l_i)}^{Tx(l_j)} < \frac{P_r(l_i)}{\gamma} - N \qquad (1)$$

Here, $l_i$ and $l_j$ are the $i^{th}$ and the $j^{th}$ link operating on the same channel, respectively. $Tx(l_i)$ and $Rx(l_i)$ are the transmitter and receiver of the link $l_i$, respectively. $P_r(l_i)$ is the receiving power at $Rx(l_i)$ due to the transmitter $Tx(l_i)$. $N$ is the background noise power. $I_{Rx(l_i)}^{Tx(l_j)}$ is the received interference power at $Rx(l_i)$ due to the transmitter $Tx(l_j)$ of link $j$ operating on the same channel. If the two-ray ground propagation model is used, $I_{Rx(l_i)}^{Tx(l_j)}$ can be easily calculated according to [11].

The current maximum admissible interference power (CMAIP) of the receiver on the link $l_i$ in the channel can be obtained from (1) as:

$$CMAIP\left[Rx(l_i)\right] = \frac{P_r(l_i)}{\gamma} - \sum_{l_j, j \neq i} I_{Rx(l_i)}^{Tx(l_j)} - N \qquad (2)$$

Then, the interference power $P_{itf}$ incurred by the new wireless link, which will be scheduled on the same channel with $l_i$, should be less than the minimum value of the CMAIP of all links on this channel. This interference can be represented as:

$$P_{itf} < \min_{l_i}\left\{CMAIP\left[Rx(l_i)\right]\right\} \qquad (3)$$

Similarly, the interference power with the new wireless link due to the links on the target channel can also be obtained. The interference power should also satisfy (1).

When any new wireless link is scheduled to use or release the channel, $P_{itf}$ should be updated.

## 4. CONTROL CHANNEL ALLOCATION ALGORITHM

The conflict graph is applied to minimize the co-channel interference in the cellular network in [12]. Then, the vertex coloring algorithm (VCA) is exploited to determine whether a set of transmissions can occur simultaneously or not. The conflict graph in WMN can be constructed as follows. Consider a graph, $G$, with vertices corresponding to mesh nodes and with edges corresponding to the single hop wireless links. The resulting conflict graph, $F$, which comes from $G$, has vertices corresponding to the edges in $G$, and has an edge between vertices in $F$ if and only if the wireless links denoted by the vertices in $F$ interfere with each other in the network topology. For example, in Figure 3-(a), there are three single-hop wireless links: BA, BC, and BD. The corresponding conflict graph should be as shown in Figure 3-(b). Obviously, if three different orthogonal channels are allocated to the vertices in Figure 3-(b), and node B has at least 3 radios tuned on different channels, there would be no interference among the 3 links. However, in practical networks, it is impossible to equip any node with an unlimited number of radios in most cases. For instance, if there are only two radios equipped on node B, that node cannot operate on three orthogonal channels simultaneously. So, its conflict graph cannot be directly applied to our system model without a constraint on the number of radios.

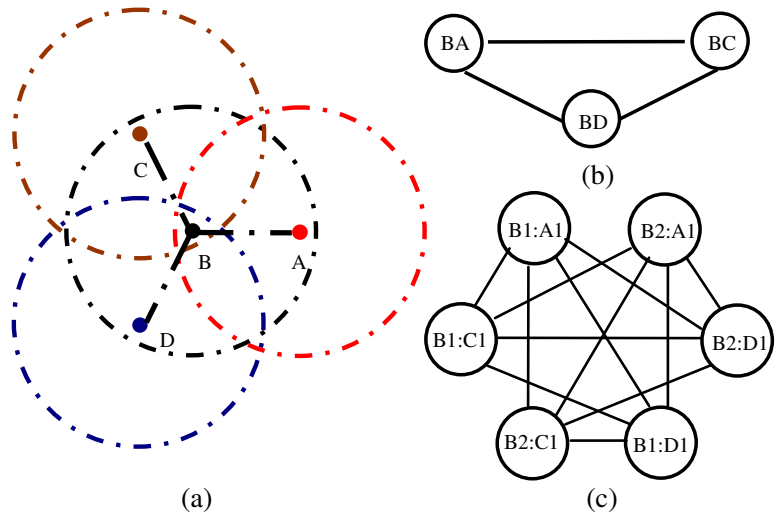

Figure 3.   (a) Network topology; (b) Conflict graph; (c) MCG

To tackle the above problem, a Multi-radio Conflict Graph (MCG) is proposed to model the interference in [6], which this paper calls the Rama scheme. Edges between radios on different nodes are represented as vertices in MCG rather than representing them between nodes in the conflict graph. Thus, the construction of MCG can be described as follows. First, each radio in the mesh network is represented as a vertex in the topology graph $G$. The edges in $G$ are between radios on different nodes. Then, the vertices in the MCG, notated as $M$, are represented as the edges in $G$. The edges between the vertices in $M$ are created in this way, i.e., two vertices in $M$ have an edge between each other if the edges in $G$ (as represented by the two vertices in $M$) interfere with each other. Figure 3-(c) is the corresponding MCG of the network topology illustrated in Figure 3-(a). The MCG comes with the constraint that nodes A, B, C, and D must be equipped with 1, 2, 1, and 1 radio(s), respectively. In Figure 3-(c), the vertex (B2:C1) represents the wireless link between the second radio on node B and the first radio on node C.

To ensure connectivity, the Rama scheme requires that at least one radio of each node tunes to the single common channel, which is also used as the single CCC, and one of data channels. Additionally, regarding coloring any vertex in MCG, the Rama scheme has a constraint that all uncolored vertices in MCG that contain any radio from the just-colored vertex must be removed. In this case, this method will lead to a considerable imbalance in load among channels and serious co-channel interference on the common channel, which in turn significantly degrades the network performance. For example, if node A in Figure 3-(a) is closest to the gateway, after allocating channel 1 and 2 to link AB and BC, respectively, all the vertices in Figure 3-(c) should be removed. In this case, link BD should be allocated a randomly selected channel without considering the interference and load according to the Rama scheme.

Unlike our method in the Rama scheme, we discard the requirement in TSC-M2MAC as well as the constraint in the Rama scheme. Instead, we avoid the scenario of removing the vertices representing the wireless links to which orthogonal channels are unallocated, from the view of the mesh node rather than the radio. For example, after allocating channels to link AB and BC, vertices like (B1:D1) and (B2:D1), which present the unassigned link BD, should not be removed.

Without loss of generality, assume that there are $N$ nodes in a WMN, and that node $K$ has $N_K$ radios, denoted by $R_K = \left(K_1, K_2, \ldots, K_{N_K}\right), K \in \left(1, \ldots N\right)$. Then, the wireless links between the radios on node $P$ and $Q$ can be denoted by

$$S_{PQ} = \left\{ Ps : Qt \middle| \left\langle Ps, Qt \right\rangle \in R_P \times R_Q \right\}, P, Q \in \left(1, 2, \ldots N\right) \qquad (4)$$

Where "×" represents the operation of the Cartesian product. Obviously, if node $P$ and node $Q$ are in transmitting range of each other, such a link exists; otherwise, it does not exist and can be denoted by $\boldsymbol{0}$. Clearly, $S_{PP} = \boldsymbol{0}$, $P \in (1, 2, \ldots, N)$. As a result, the wireless links in the mesh network can be represented by a matrix:

$$S = [S_{PQ}]_{N \times N} = \begin{bmatrix} 0 & S_{12} & \cdots & S_{1N} \\ S_{21} & 0 & \cdots & S_{2N} \\ \vdots & \vdots & \ddots & \vdots \\ S_{N1} & S_{N2} & \cdots & 0 \end{bmatrix} \qquad (5)$$

In order to solve the problem in the Rama scheme and to ensure only one channel is allocated to each radio, a significant constraint should be imposed in CCAA when using BFS-based VCA to color the MCG: once any vertex (Ps,Qt) is colored, then all the vertices corresponding to the wireless links in $S_{PQ}$ should be completely removed. In this way, there could exist a scenario whereby there are not enough orthogonal channels to be allocated to the links in the interference range, or there may be an insufficient number of radios on the node. In this case, some of these links must share channels, although interference will occur when transmitting and/or receiving simultaneously. Such a scenario is inevitable when the scale of a network is large and the density of nodes is intensive. To minimize co-channel interference in such a scenario and to balance the load on each channel, the sharing channel should be selected according to the hop counts of the node from the gateway and the accessible conditions for new wireless links. The links with less hop counts should be assigned higher priority to share channels that have a smaller number of links, considering the traffic pattern in WMN. So, the uncolored vertices encountering this scenario should try to share channels that have links of the same priority. In this way, the connectivity of the network is ensured and the topology remains unchanged. The CCAA is demonstrated in Table 1.

Table 1.  Control Channel Allocation Algorithm (CCAA)

| |
|---|
| Initialization: $V=\{v \mid v \in MCG\}$; $C=\{all \quad channels\}$; $h=1$; |
| **while** $h <= maximumHopCount$ **do** |
| $\quad$ $Queue=\{v \mid v \in V$ and $notVisisted(v)$ and $hopCount(v)=h\}$ |
| $\quad$ **while** $size(Queue)>0$ **do** |
| $\quad\quad$ $v_{current}=popHead(Queue)$ |
| $\quad\quad$ **if** $visited(v_{current})=True$ **then** |
| $\quad\quad\quad$ **continue** |
| $\quad\quad$ **end if** |
| $\quad\quad$ $visit(v_{current})$ |
| $\quad\quad$ $V_{neighbor}=\{v \mid v \in V$ and $edge(v, v_{current})$ exists $\}$ |
| $\quad\quad$ $V_{temp}=\{v \mid v \in V_{neighbor}$ and $v$ has been visited$\}$ |
| $V^{'}_{temp}=\{v\|v\in V_{neighbor}$ and $v$ has not been visited$\}$ |
| $C_{available}=\{c\|c\in C$ and $c$ does not conflict with the channels that had been assigned to $v_i\in V_{temp}\}$ |
| **while** $V_{temp}$ is not empty **do** |
|     **if** $C_{available}$ is not empty **then** |
|         Assign a randomly selected channel from $C_{available}$ to $v_m$, $\{v_m\in V^{'}_{temp}\}$ |
|         Remove all the vertices corresponding to $S_{PQ}$ from MCG with $(Ps:Qt)$ denote $V_m$ |
|         Update $V_{temp}$, $V^{'}_{temp}$ and $C_{available}$ |
|     **else** |
|         Assign channel to $v_m$, $\{v_m\in V^{'}_{temp}\}$ according to the admissible condition |
|         Remove all the vertices corresponding to $S_{PQ}$ from MCG with $(Ps:Qt)$ denote $V_m$ |
|         Update $V_{temp}$, $V^{'}_{temp}$ and $C_{available}$ |
|     **end if** |
|     **end while** |
|   **end while** |
|   $h=h+1$ |
| **end while** |

In contrast to the single CCC in the Rama scheme, CCAA uses all available channels as control channels. This type of allocation can indicate which radio on each node should tune to which channel to exchange control messages. It also requires few system resources as long as the executing frequency is not large, which is crucial for the gateway. If the gateway is overloaded, then it will not have enough resources to deal with functions like forwarding packets between the mesh network and the outside Internet.

In addition to these advantages, CCAA perfectly solves the broadcast problem in multiple control channels because of two factors. First, it ensures the connectivity of the nodes in WMN. Second, the CRUS, defined in section 5.1, is shared by all radios on each node and can therefore ensure that all radios are accessible as long as one of them on the node captures the broadcast packet. In these ways, CCAA can reduce the overhead caused by transmitting duplicated broadcast packets in the previous broadcast mechanism.

# 5. DYNAMIC DATA CHANNEL ALLOCATION

## 5.1. Criterion determining the channel and radio utilization structure (CRUS)

In TSC-M2MAC, the available channels are categorized into two groups: idle channels and busy channels. The priority for every idle channel is the same, and they are selected randomly. The priority of idle channels is higher than that of busy channels. For busy channels, the priority for each channel is determined by CMAIP, as defined in (2). The more interference a channel has, the lower its priority is. If CMAIP is the same, then the channel with the lowest traffic load has the highest priority. CRUS is a data structure consisting mainly of three items: node ID, a list of sorted available channels in terms of priority, and a map indicating the radios' channel utilization on the node. Every node maintains its own CRUS.

## 5.2. Dynamic channel allocation for data transmission

In TSC-M2MAC, a REQ/ACK/RES mechanism is used to allocate the channels for data transmission. All control messages are exchanged over the CCAA's resulting channels. The second coordination stage contains five steps:

Step 1: Once a beacon comes, if a source node (node B) has data pending for a destination (node C), B should check its $CRUS_B$ to find whether a radio and an available channel can be used for data transmission. If so, B waits for $T_{DIFS}$ and a random exponential backoff value, and then transmits a packet $REQ(CRUS_B)$ to C and broadcasts it to B's neighbors. If node B does not have data pending, then B must wait until the next beacon comes.

Step 2: Once node C receives the packet $REQ(CRUS_B)$, C has to check whether it has idle radios. If it does not, then C sends an acknowledgment message ACK(INVALID) to B and broadcast it to C's neighbors to inform them that all radios on C are busy. If it does have idle radio, then C needs to select a channel for data transmission according to $CRUS_B$ and $CRUS_C$. If the intersection between $CRUS_B$ and $CRUS_C$ is empty, node C sends an ACK(NULL) message to tell B that no channel is available for data transmission, and also broadcasts it to C's neighbors. It is clear that whether a receiver can decode the data correctly or not depends only on the SNIR at the receiver. So, it has no relation to $CRUS_B$ on the condition that a common channel can be found in the intersection between $CRUS_B$ and $CRUS_C$. Therefore, if the intersection is not empty, then the channel to be selected should be satisfied with one condition: it must have the highest priority in the intersection between $CRUS_B$ and $CRUS_C$, according to the sorted priority of $CRUS_C$ rather than $CRUS_B$. If the channel selected by C is $CH_k$, then the C will send the $ACK(CH_K)$ to B and broadcast it to the neighbors of C.

Step 3: Once B receives the ACK message, it will check whether the information carried by ACK is NULL, INVALID, or $CH_K$. If it is NULL or INVALID, B cancels the negotiation process and goes to step 1. If it is $CH_K$, then B will check if this channel can still be used. If that is the case, B updates its $CRUS_B$ and then broadcasts a reservation message $RES(CH_k)$ across the network. When C receives this reservation message, it updates its $CRUS_C$.

Step 4: After the exchanging process for control messages is finished, both B and C have to wait for the coming of the data pilot.

Step 5: All the nodes start to transmit data to their destinations when the data pilot arrives, and they continue to transmit data until the next beacon comes.

One must note that all the control messages, REQ, ACK, and RES, are transmitted over the allocated control channels from the first coordination stage. At this stage, CCAA is able to indicate the mapping relationship for which radio should be tuned to which control channel. Obviously, the hidden node problem can be alleviated in TSC-M2MAC. For example, there are three nodes, A, B and C. The pairing of A and B and the pairing of B and C are within the transmission range of each other, respectively. But, the pairing of A and C are beyond the transmission range of each other. If a channel is assigned to the link between A and B, it would not be selected from $CRUS_B$ when C negotiates the transmission channel with B.

## 6. AAA FOR CNA SUB-INTERVAL

In TSC-M2MAC, the fixed size of the CNA sub-interval greatly impacts the network's performance. Performance mainly depends on the NCTF scheduled on each orthogonal channel. If the NCTF is small, it will require a short time for negotiation and will need a small size of CNA sub-interval. Otherwise, the collision probability of control messages will become high, and the negotiation will take more time, which requires a large size of CNA sub-interval. So, its size should be adaptively adjusted according to the practical NCTF. On the other hand, the CNA size should not be infinitely increased, since enough time should be left for data transmission. So, a threshold should be set to limit the increase of the CNA size. If the CNA size has been equal to the threshold, those communication pairs that have not yet finished negotiation will be denied. Such a threshold heavily depends on the network's topology and its traffic patterns, so the threshold should be determined according to the practical case.

In addition, the efficiency of AAA heavily depends on whether the scheduling of traffic flows on the available channels is balanced or not, because the same time division method is used for all the channels. For example, if one channel was overloaded whereas others were under-loaded, there would be no ideal dynamic adjustment scheme available. Consequently, in TSC-M2MAC, load balancing should be taken into consideration at both coordination stages.

To cope with the instability of the network, the size of the CNA sub-interval can only be increased or decreased a one-step size. In the meantime, when the size is decreased, a margin of one step size should be reserved so that it can be used to avoid CNA oscillation and also to ensure that the size of the CNA sub-interval is able to converge into a stable state. The AAA of the CNA sub-interval is in Table 3, and the related notations in the algorithm are in Table 2.

Table 2.    Notations in AAA for CNA Sub-interval

| Notations | Meaning |
|---|---|
| $CNA$ | The size of CNA sub-interval |
| $CNA_{min}$ | The minimum size of CNA sub-interval |
| $CNA_{max}$ | The maximum size of CNA sub-interval |
| $Step$ | The length of time that the CNA sub-interval takes to increase or decrease one time |
| $AccIdleT_i$ | The accumulated idle time of channel $i$ |
| $Threshold_{adj}$ | The threshold to adjust the size of CNA sub-interval |

Table 3.    AAA for CNA Sub-interval

| |
|---|
| Initialization: All the radios begin with $CNA= CNA_{min}$ |
| When time approaches the end of $CNA$ |
| **if** $minimum(AccIdleT_i) <= Threshold_{adj}$ **then** |
| **if** $CNA_{max} <= CNA + Step$ **then** |
| $CNA = CNA_{max}$ |
| **else** increase $CNA$ by step |
| **end if** |
| **else if** $minimum(AccIdleT_i) > CNA + Step$ **then** |
| **if** $CNA - Step < CNA_{min}$ **then** |
| $CNA = CNA_{min}$ |
| **else** decrease $CNA$ by $Step$ |
| **end if** |
| **else** |
| remain $CNA$ unchangeable |
| **end if** |
| Repeat the above process until all the events are scheduled |

In TSC-M2MAC, the gateway executes such an adjustment algorithm. The algorithm may occupy much system resource in order to time synchronize and coordinate all nodes. So, similar to CCAA, it should also be executed for a relatively long period rather than every time interval if the traffic pattern varies slowly.

# 7. POWER SAVING MECHANISM IN TSC-M2MAC

Energy consumption is a key challenge that requires an urgent solution due to current concerns about cost, the environment, the limitation of non-renewable resources, etc. Various types of measures can be taken to tackle this problem. However, this paper will focus only on how to intelligently let radios on each node transit among the state space to save power without adversely affecting the performance of the network.

## 7.1. Problem formulation

A radio usually has the following four states: busy (transmitting or receiving), idle (ready to transmit or receive data), doze (cannot transmit or receive and thus consuming little energy) and off state (consuming no energy). Table 4 outlines the energy consumption of two typical commercial radios [13]-[15]. It shows that energy can be saved by switching a radio from idle state to doze state if it has no packets to exchange. When they are needed, the radios in doze state should be awaked. The switching process among state spaces can be described by a discrete event dynamic system. In the process, the transition is driven by different events that are arriving at the system. Each event is triggered by the timeout mechanism set on each radio. For each arrived event, corresponding decisions should be made according to the given rules. Consequently, energy consumption in TSC-M2MAC can be formulated as an optimization problem: how to make a decision for each radio when a new event comes such that the total energy consumption is minimized subject to the constraint that the requirement for system performance is satisfied. The main assumption in the proposed PSM is that the time spent on transition from one state to another is far less than the time interval used in the proposed TSC-M2MAC protocol. Otherwise, such a mechanism will become meaningless.

Table 4.　Power Consumption (unit: watt)

| Type | Tx | Rx | idle | doze |
|------|-----|-----|------|-------|
| Laucent WaveLAN | 1.65 | 1.4 | 1.15 | 0.045 |
| Cisco AIR-PCM 350 | 1.88 | 1.3 | 1.08 | 0.045 |

The off state is not considered in the proposed PSM, since it is different to awake. As a result, the state space contains three states: S={busy, idle, doze}. The event set can be represented as E={Et, t≥0}. The decision space can be denoted by D={D1, D2, …, Dn}. Below, the state evolution process is illustrated by Figure 4.

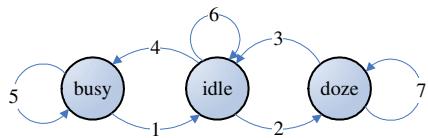

Figure 4.　The state evolution process of each radio

## 7.2. Decisions for power saving

As far as the radios in our scheme are concerned, there are only 7 decisions (denoted by the numbers 1 to 7 shown in Figure 4) no matter which event comes. Among them, only decision 2, 3, and 7 are related to power saving. The mapping relationship between events and the 3 decisions can be described as follows:

Decision 2: this decision will be made if two conditions are satisfied. First, the period between the epoch when one radio finishes transmitting or receiving messages and the following time indicator (beacon or pilot) is longer than the time spent on the sleeping process (transition from idle to doze) and the waking process (inverse transition). Second, the saved energy by means of making the decision is larger than the additional energy spent on the sleeping and waking processes.

Decision 3: if the following time indicator is a beacon, then such a decision should be made at the moment of a period spent on the waking process before the next beacon comes. If the following time indicator is a pilot and the radio also has data to exchange, then this decision will be made at the moment of a period spent on the waking process before the next pilot comes.

Decision 7: this decision will be made if the radio finishes the coordination during the CNA sub-interval and has no data to exchange with other radios.

Decision 3 indicates that the radio can transit from doze state to idle state through the timing mechanism, which does not require any additional control information. The timing mechanism can also be used in decision 7.

As a result, as long as events corresponding to the three decisions listed above come, the resulting action from the corresponding decision will be taken. Consequently, these economical decisions can decrease the total energy consumption.

## 8. SIMULATION-BASED PERFORMANCE EVALUATION

### 8.1. Performance of the control channel allocation algorithm (CCAA)

From section 4, it can be seen that the CCAA mainly depends on the network topology, NOC, the traffic pattern, etc. It is difficult to completely evaluate the algorithm from all aspects. So, we use the case study method to simplify the simulation. We assume that there is one gateway, 3 available orthogonal channels, 10 and 100 nodes with one and two hops respectively from the gateway. We use the throughput fairness index defined in [16] to measure the load balancing, i.e., throughput balancing among available channels. Obviously, the higher the fairness index is, the better the performance is, in terms of load balancing. We compare the proposed CCAA with the Rama scheme using different randomly generated topologies.

Table 5.   Fairness Index with Randomly Generated Topologies

| Topology | 1st | 2nd | 3rd | 4th | 5th |
|---|---|---|---|---|---|
| CCAA | 0.72 | 0.68 | 0.86 | 0.83 | 0.79 |
| Rama scheme | 0.47 | 0.44 | 0.52 | 0.54 | 0.63 |

Table 5 shows that the proposed CCAA outperforms the Rama scheme with respect to fairness of throughput among available channels. The reason why the fairness for CCAA is less than 1 is that the links closer to the gateway have higher priority on allocating channels than those far from the gateway. Load balancing plays an important role in the proposed TSC-M2MAC, since the channels with the heaviest load determine the size of the CNA sub-interval, and this determination significantly impacts the network's overall performance.

### 8.2. Performance of TSC-M2MAC

### 8.2.1. Efficiency of TSC-M2MAC

The performance of TSC-M2MAC is characterized by efficiency and stability. We use average aggregated throughput and end-to-end packet delay to evaluate its efficiency, and we use packet loss rates to evaluate its stability, similar to [17]-[19]. Although few similar protocols with TSC-M2MAC exist in the previous literatures, measures have been taken to decrease the contentions in the single CCC [20]-[23]. To evaluate the efficiency of the proposed TSC-M2MAC, we compare it with three schemes: 1) the "pure" IEEE 802.11 MAC protocol with a single CCC and a single data channel; 2) the Rama scheme, proposed in [6], with a single CCC and multiple data channels; 3) the enhanced Rama (en_Rama) scheme with an additional mechanism of the adaptive contention window proposed in [20]. We apply the same number of orthogonal channels in all of our comparisons of the TSC-M2MAC, Rama, and en_Rama. Additionally, to evaluate the stability of the TSC-M2MAC, we observe the packet loss rate when the aggregated transmission rate is increased from a small value to a value that is large enough to saturate the network. We perform the simulations using the ns2 simulator with CMU wireless extensions [24]. The number of radios equipped on each node is a uniformly distributed integral number with two bounds: the lower bound is 1 and the upper bound corresponds to the number of orthogonal channels. The handover time for a radio from one channel to another is assumed to be 224 $\mu s$, and the time interval is assumed to be 100ms (refer to [1][9]). In this part, we set the size of the CNA sub-interval as a fixed value (20% of the time interval) and we devote the rest of the time interval for DT. We will simulate the AAA of the CNA sub-interval in part 8.3. The following common parameters are used in the simulation. The radio power and threshold levels are set such that the transmission range and the carrier sensing range are 250m and 550m, respectively. The bandwidth of each channel (including the CCC and data channels) is 1Mbps. We use a wired-cum-wireless topology on a 1500m-by-1500m area with one gateway, one wired node, and 23 randomly positioned wireless nodes in the wireless domain of the gateway. The bandwidth, delay, and propagation model of the duplex-wired link between the wired node and the gateway are 100Mbps, 1ms and drop tail, respectively. The two-ray ground reflection model is used to model the wireless propagation. There are 13 CBR flows with source-destination pairs. Three of the CBR flows are destined to the wired node via the gateway, and the others are random selected pairs. Each simulation is performed long enough to saturate the network. Each data point in the plot has the averaged value of 30 runs with different topologies. The error bars show 95% confidence intervals for the difference of each run.

We plotted the average aggregated throughput and the average end-to-end packet delay in Figure 5 and Figure 6, respectively, for the TSC-M2MAC, pure 802.11, the Rama schemes, and the enhanced Rama scheme with an adaptive contention window. The abbreviations in Figure 5 and Figure 6 are as follows: 1ch_802_11 denotes the 802.11 scheme; 2ch_TSC-M2MAC, 4ch_TSC-M2MAC denote the TSC-M2MAC with 2, 4 orthogonal channels, respectively; 2ch_Rama, 4ch_Rama denote the Rama scheme with 2, 4 orthogonal channels, respectively; and 2ch_en_Rama, 4ch_en_Rama denote the enhanced Rama scheme with 2, 4 orthogonal channels, respectively. Figure 5 depicts that: 1) the average aggregated throughputs increase as the number of orthogonal channels increases in TSC-M2MAC, Rama, and enhanced Rama schemes for any specific packet size; however, the throughputs achieved by TSC-M2MAC using 2 and 4 channels are always higher than the throughputs achieved by Rama and enhanced Rama schemes using the same number of channels, although enhanced Rama can achieve higher throughput than Rama, which shows the efficiency of the utilization of multiple control channels; 2) the increasing speed of the throughputs becomes moderate with the increase in packet size when the packet size is less than 1100bytes; and the throughputs become nearly stable when the packet size is larger than 1100bytes; 3) TSC-M2MAC, Rama, and enhanced Rama schemes can achieve much higher throughputs than 802.11, and this shows the advantage of using multiple orthogonal channels; 4) the improvement on throughput over 802.11 MAC is still less than $k$ times when $k$ orthogonal channels are used in TSC-M2MAC, which may be caused by the fact that the REQ/ACK/RES is more complex and requires more coordination time than the RTS/CTS used in 802.11. In short, utilizing multiple orthogonal channels for data transmission within a single control channel can improve higher throughput over 802.11. Also, the TSC-M2MAC with multiple control channels outperforms the Rama and the enhanced schemes with a single common control channel in terms of throughput when the same number of in-band orthogonal channels is used.

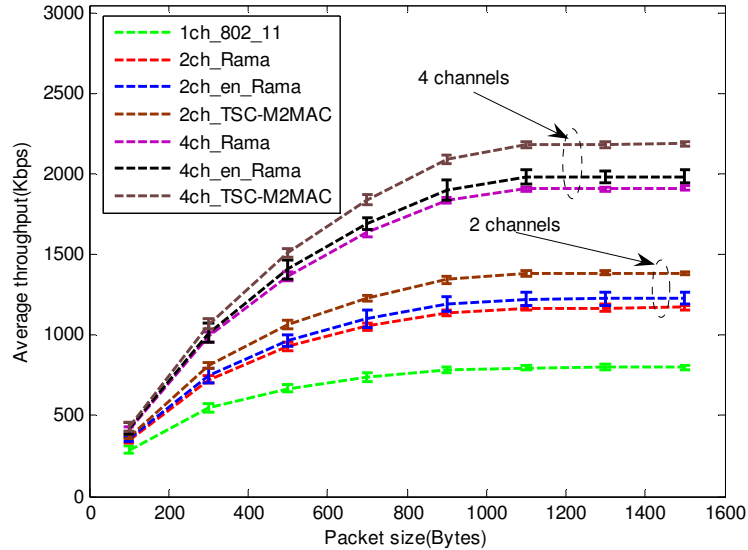

Figure 5.   Average aggregated throughput with different packet sizes

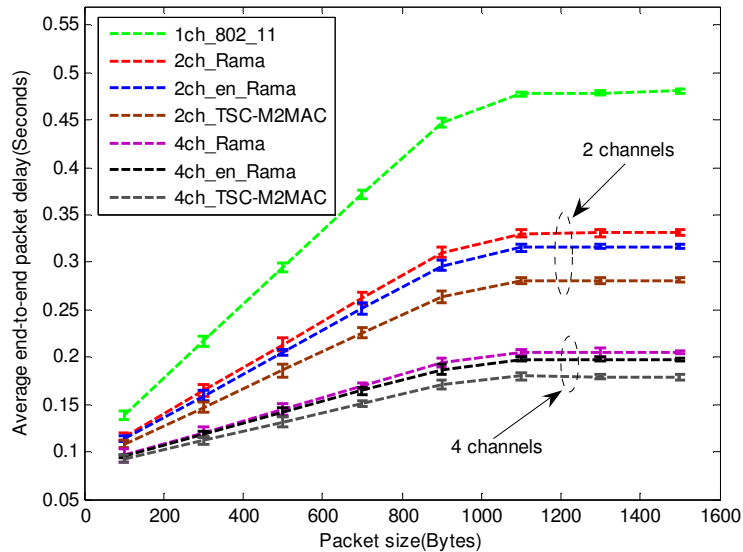

Figure 6.   End-to-end packet delay with different packet sizes

Figure 6 shows the comparison of the average end-to-end packet delays in the TSC-M2MAC, Rama, enhanced Rama, and 802.11 schemes as the packet size increases. The results indicate that: 1) the delay decreases as the number of orthogonal channels increases when it has the same packet size; 2) the delays increase with the increase in packet size when the packet size is smaller than 1100 bytes, and it becomes nearly stable when it is larger than 1100 bytes in all three schemes; 3) the increasing speed of delays becomes moderate with the increasing of packet sizes for any specific number of channels; 4) the delays in both TSC-M2MAC and Rama schemes are lower than those in 802.11; 5) the delay in TSC-M2MAC is always lower than the delay in Rama and in the enhanced Rama schemes with the same orthogonal channels, although enhanced Rama can achieve lower delay than the Rama scheme. In short, the utilization of multiple control channels can achieve lower end-to-end packet delay than that of a single, dedicated common control channel.

Figure 5 and Figure 6 show that TSC-M2MAC can achieve the highest throughput and the lowest end-to-end packet delay among the four schemes. On the opposite end of the spectrum, the 802.11 scheme has the worst performance in terms of throughput and delay. It can be seen that the proposed TSC-M2MAC can further improve the throughput and reduce end-to-end packet delay better than the Rama and enhanced Rama schemes using the same number of orthogonal channels. This ability shows the advantage of using multiple control channels over a single common control channel, and the TSC-M2MAC does not face the bottleneck that arises in the single dedicated CCC. Notice the fact that the existing 802.11a, 802.11b/g can provide 13 and 3 available orthogonal channels, respectively, and the widely used 802.11 MAC wastes a lot of spectrum resource, resulting in performance degradation; it is therefore the worst one among the 4 schemes compared here.

### 8.2.2. Stability of TSC-M2MAC

In our stability evaluation, we use the same parameters as we used in section IV-A. Similar to [17]-[19], packet loss rate is selected as the evaluation metric. We observe the packet loss rate when the transmission rate increases with a specific packet size (210 bytes). We use the packet loss rate of the pure IEEE 802.11 as the baseline. The packet loss rate appears in Figure 7.

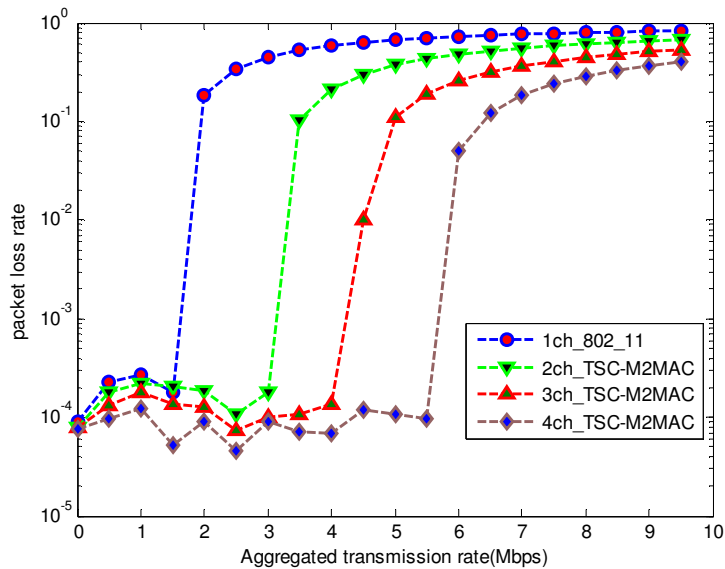

Figure 7.    Packet loss rate with a given packet size (210 bytes)

Figure 7 shows that the packet loss rate in both pure 802.11 and TD-MAMAC fluctuate before the network is saturated. Recall that the bandwidth of each channel is assumed to be the same, but the corresponding aggregated transmission rates for saturation with a different number of channels appear differently in Figure 7. This appearance also indicates that both pure 802.11 and the proposed TSC-M2MAC will become stable when the network saturates. In addition, Figure 7 shows that the packet loss rate decreases with the increase in the number of orthogonal channels. Since the loss rate is defined as the ratio between the number of packets unsuccessfully accepted by the destinations and the total transmitted packets, all 4 curves approach 100% when the transmission rate increases.

### 8.3. Evaluation of AAA for CNA sub-interval

As mentioned before, the contention time for channel negotiation has a dominating impact on the performance of TSC-M2MAC. Thus, at first, we simulate the consumed contention time as NCTF increases, as shown in Table 6. Table 6 indicates that the average contention time increases sharply as NCTF increases, because the exponential backoff mechanism begins when collision happens.

Table 6. Averaged contention time vs. Number of concurrent traffic flows

| NCTF | 10 | 25 | 40 | 55 | 70 | 85 | 100 |
|---|---|---|---|---|---|---|---|
| Contention time (ms) | 3.78 | 7.31 | 12.61 | 23.07 | 38.44 | 57.81 | 80.62 |

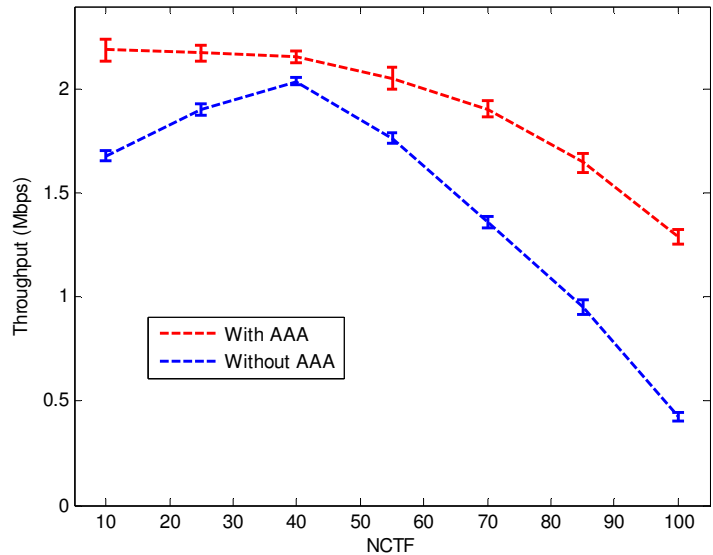

Figure 8. Impact of AAA on the throughput

We also apply a case study method to evaluate the efficiency of the proposed AAA of the CNA sub-interval, since the measurement varies with many factors, such as NOC, NCTF, etc. We take 4 available channels as a special case and measure the aggregated throughput as the NCTF increases. In the simulation, we set the threshold for the CNA sub-interval at 50ms, which is half of the time interval. We compare the achieved throughput by exploiting AAA of the CNA sub-interval, with that employing a fixed CNA sub-interval to 20ms. Figure 8 shows that the throughput achieved using AAA is always higher than that using fixed CNA size. It also shows that the throughput with AAA drops slowly and sharply when NCTF is less and more than 40, respectively. This is because the time available for data transmission becomes shorter and shorter as NCTF increases. However, with the fixed CNA sub-interval, the throughput increases when NCTF is less than 40 and drops rapidly when NCTF is more than 40. This is because the CNA sub-interval is so large that it becomes wasteful when there are few concurrent flows, and then it becomes too small for negotiation, resulting in service denial for the flows not finishing negotiation when NCTF becomes large. It also shows that when there are 100 concurrent flows, the achieved throughput for the fixed CNA sub-interval is much less than that for the AAA case. The reason is that too many flows with large contention time are denied for the fixed sub-interval case, while for AAA, time for negotiation can occupy as much as half of the time interval, which is much larger than the fixed CNA sub-interval. Thus, the adaptive adjustment of CNA outperforms the fixed case in terms of aggregated throughput.

### 8.4. Performance of the power saving mechanism

In order to evaluate the performance of the proposed PSM with a simple simulation, we must establish a few main assumptions. First, the buffer on each node is infinite and there is always data to transmit. Second, the sequence of the packet for transmission follows the rule of FCFS (First Come, First Serve). Third, the transmitted data for each request follows an exponential distribution with a mean equal to 100M bits. Fourth, the powers of the radio on each state are referred to the interface card of *Laucent WaveLAN*, and the data transmission rate on each channel is 1Mbps. Each point in the plot is the average of the achieved results of our simulation, which made 30 runs at different seeds. In our simulation, we take the scenario with two orthogonal channels as an example to demonstrate the performance of the proposed PSM. It is easy to generalize such a mechanism to other scenarios with variant numbers of orthogonal channels. The consumed energy with and without PSM is compared in our simulation.

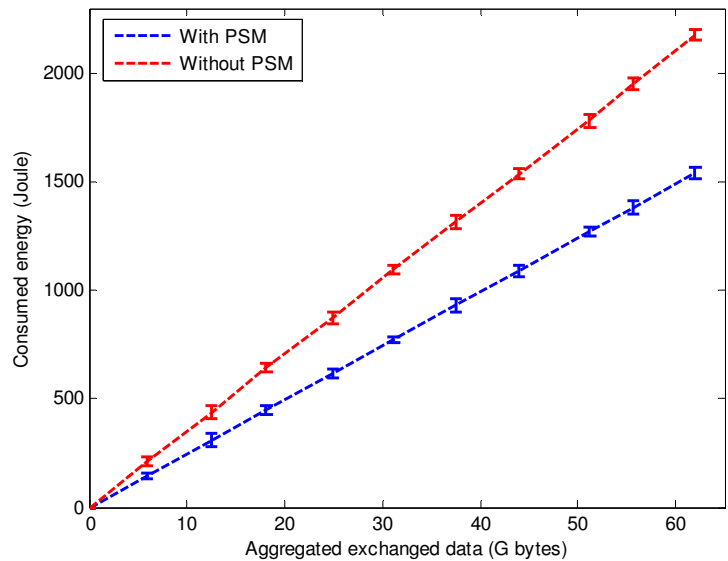

Figure 9.   Performance of the power saving mechanism (PSM)

The comparison of energy consumption with and without PSM is shown in Figure 9. We have calculated that the energy saved in the case with PSM is approximately 29% compared to the energy in the case without PSM. It can therefore be expected that the system with PSM can save more energy during the period of less requests from clients, e.g., the period near dawn for the multimedia Video on Demand (VoD) system.

## 9. CONCLUSIONS

Instead of using a single dedicated Common Control Channel (CCC), the proposed TSC-M2MAC designates all available channels as control channels on the Channel Negotiation and Allocation (CNA) sub-interval and the data channels on the Data Transmission (DT) sub-interval in a time division manner through a two-stage coordination. At the first coordination stage, a Multi-radio Conflict Graph (MCG) is used to model the co-channel interference, and the Breadth-First-Search (BFS)-based Vertex Coloring Algorithm (VCA) is used to realize an intelligent control channel allocation. As a result, the co-channel interference decreases. At the second stage, a REQ/ACK /RES mechanism is proposed to implement the dynamical channel allocation for data transmission. The hidden node problem is successfully alleviated by using a Channel and Radio Utilization Structure (CRUS). Then, the problems that arise from the variation of Number of Concurrent Traffic Flows (NCTF) are mitigated using the Adaptive Adjustment Algorithm (AAA) for the CNA sub-interval. Simulation results show that TSC-M2MAC can achieve load balancing among multiple channels; as the NCTF increases, it can achieve higher throughput and lower end-to-end packet delay than conventional methods. Also, TSC-M2MAC converge into stability when the network saturates; finally, the effectiveness of the Power Saving Mechanism (PSM) proposed specially for TSC-M2MAC has also been verified, because it helps to save 29% of the total power consumption.

**Authors' Short Biography**

**Bingxuan Zhao**

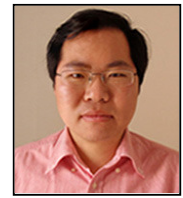

Bingxuan Zhao was born in Puyang, Henan province, China, on the 20th November, 1981. He received the B.E. degree with the outstanding graduate in electronic information engineering from Jilin University, M.E. degree in network communication system and control from University of Science and Technology of China, in 2005 and 2008, respectively. Currently he is pursuing the Ph.D. degree in the graduate school of global Information and telecommunication studies, Waseda University, Japan. His research interests include cognitive radio and wireless mesh networks.

**Shigeru Shimamoto**

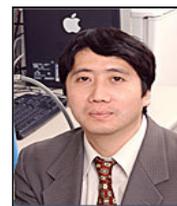

Shigeru Shimamoto was born in Mie, Japan in 1963. He received the B.E and M.E. degrees in communications engineering from the University of Electro Communication, Tokyo, Japan, in 1985 and 1987, respectively. He received the Ph. D. degree from Tohoku University, Japan in 1992. From April 1987 to March 1991, he joined NEC Corporation. From April 1991 to September 1992, he was an Assistant Professor in the University of Electro Communications, Tokyo, Japan. He has been an Assistant Professor in the Gunma University from October 1992 to December 1993. Since January 1994 to March 2000, he has been an Associate Professor in Department of Computer Science, Faculty of Engineering, Gunma University, Gunma, Japan. Since April 2002, he has been a Professor at GITS, Waseda University. In 2008, he also served as a visiting professor at Stanford University, USA. His main fields of research interest include wireless mesh networks, sensor networks, satellite and mobile communications, optical wireless, Ad-hoc networks and body area network.